\documentclass[%
 reprint,
 amsmath,amssymb,
aps,
prb,
]{revtex4-1}

\usepackage{graphicx}% Include figure files
\usepackage{dcolumn}% Align table columns on decimal point
\usepackage{bm}% bold math
\usepackage[caption=false]{subfig}
\begin{document}

%\preprint{APS/123-QED}

\title{Pressure dependent intermediate valence behavior in YbNiGa$_{4}$ and YbNiIn$_{4}$}% Force line breaks with \\

\author{Z. E. Brubaker$^{1,2}$, R. L. Stillwell$^{2}$, P. Chow$^{3}$, Y. Xiao$^{3}$, C. Kenney-Benson$^{3}$, R. Ferry$^{3}$, D. Popov$^{3}$, S. B. Donald$^{2}$, P. S{\"o}derlind$^{2}$, D. J. Campbell$^{4}$, J. Paglione$^{4}$, K. Huang$^{5}$, R. E. Baumbach$^{5}$, R. J. Zieve$^{1}$ and J. R. Jeffries$^{2}$\\
$^{1}$ Physics Department, University of California, Davis, California, USA\\
$^{2}$ Lawrence Livermore National Laboratory, Livermore, California 94550, USA\\
$^{3}$ HPCAT, Geophysical Laboratory, Carnegie Institute of Washington, Argonne National  
Laboratory, Argonne, Illinois 60439, USA\\
$^{4}$ Center for Nanophysics and Advanced Materials, Department of Physics, University of Maryland, College Park, Maryland, 20742, USA\\
$^{5}$ National High Magnetic Field Laboratory, Florida State University, Tallahassee, FL 32313, USA\\
}
\date{\today}% It is always \today, today,
                     
\begin{abstract}
We report a comprehensive structural and valence study of the intermediate valent materials YbNiGa$_{4}$ and YbNiIn$_{4}$ under pressures up to 60 GPa. YbNiGa$_{4}$ undergoes a smooth volume contraction and shows steady increase in Yb-valence with pressure, though the Yb-valence reaches saturation around 25 GPa. In YbNiIn$_{4}$, a change in pressure dependence of the volume and a peak in Yb-valence suggest a pressure induced electronic topological transition occurs around 10-14 GPa. In the pressure region where YbNiIn$_{4}$ and YbNiGa$_{4}$ possess similar Yb-Yb spacings the Yb-valence reveals a precipitous drop. This drop is not captured by density-functional-theory calculations and implies that both the lattice degrees of freedom and the chemical environment play an important role in establishing the valence of Yb.

%\begin{description}
%\end{description}
\end{abstract}

\pacs{Valid PACS appear here}% PACS, the Physics and Astronomy
%\maketitle

%\tableofcontents
\maketitle
\section{\label{sec:level1}Introduction}

Strongly correlated rare-earth materials have been heavily studied due to the exotic physical properties that they exhibit. Many of the rare earths form compounds in intermediate valence states, which will naturally dictate the magnetic properties of these materials, and which can readily be tuned via the application of pressure. While in Ce-compounds, pressure generally favors a non-magnetic Ce$^{4+}$ state, in Yb-compounds pressure favors the magnetic Yb$^{3+}$ state.\cite{hfmagnet} The strong electron correlation in rare-earth bearing materials originates from 4f electrons that at normal conditions are localized on the atom. However, if the rare-earth atoms are sufficiently close, due to application of pressure or within a suitable crystal structure, they can interact, displaying behavior that adopts delocalized character. Thus, by choosing appropriate rare-earth compounds, intermediate valence behavior (degree of localization) can be achieved and tuned via the application of pressure. Amongst Yb-based compounds, one of the only known superconductors, ${\beta}$-YbAlB$_{4}$, shows valence fluctuations at 20K with an effective valence of n=2.75, which has sparked interest in better understanding the relationship between intermediate valence behavior, quantum criticality, and superconductivity. \cite{YBSC,YBSCval} The effect of the Yb-valence state on magnetic properties has also been studied in YbNi$_{3}$Ga$_{9}$, which forms a non-magnetic state at low temperature with an effective valence of n=2.6 under ambient conditions.\cite{139} With the application of pressure, the valence is increased to 2.9, allowing a magnetic ground state to develop.

Because the rare-earth valence plays such a crucial role in determining magnetic properties, it is imperative to understand both its cause and how to tailor materials to exhibit the desired valence configuration. With this goal in mind, we have performed density-functional theory (DFT) calculations as well as a comprehensive structural and valence study of the orthorhombic YbNiGa$_{4}$ and YbNiIn$_{4}$ system under pressure at room temperature. Previous work reported a valence of n=2.40 in YbNiIn$_{4}$, n=2.48 in YbNiGa$_{4}$ and n=2.66 in YbNiAl$_{4}$, showing a general trend of increasing Yb-valence with decreasing size of the group IIIb ions.\cite{YNAval,YNI,YNG} The Yb valence in YbNiGa$_{4}$ was measured up to a pressure of 25 GPa, revealing a steady increase in valence up to a maximum value of n=2.7.\cite{YNG} 

Our DFT calculations are consistent with the overall trend of the Yb-valence at ambient pressure in this system, but suggest a stronger dependence on interatomic spacing. In order to determine if the valence is a simple function of interatomic spacing, we have determined the Yb-valence in YbNiGa$_{4}$ and YbNiIn$_{4}$ under pressures up to 40 GPa. Rather than being a continuous function of atomic volume or lattice spacing, we find the Yb-valence is sensitive to both the lattice degrees of freedom as well as the chemical environment. DFT does well to reproduce the pressure dependence of the In compound up to a pressure that generates lattice spacings comparable to those of the Ga variant at ambient pressure. However, the substitution of Ga for In results in a precipitous drop in valence at fixed lattice size, an effect not captured by our DFT calculations, and one that implies a more prominent role for the 4f-hybridization with specific p-states than might be conventionally expected.

\section{\label{sec:level1}Experimental and Theoretical Methods}

Polycrystalline samples with nominal compositions of YbNiGa$_{4}$ and YbNiIn$_{4}$ were grown via arc-melting in argon atmosphere. Due to the low boiling temperature of ytterbium, a 5\% excess of ytterbium was necessary to account for the mass loss during melting. Each sample was melted and flipped 6 times. Samples were subsequently annealed at 625$^\circ$C for 10 days, and powder x-ray diffraction (PXRD) measurements were performed both before and after the annealing procedure. There was no detectable mass loss during the annealing process. PXRD analysis indicated phase purities $>$96\% for YbNiGa$_{4}$ and YbNiIn$_{4}$, with YbGa$_{2}$ and YbIn$_{3}$ being the main impurity phases, along with $<$1\% of Yb$_{2}$O$_{3}$. The Yb-compounds order in the cmcm structure, with Yb and Ni occupying the 4c Wyckoff position, and the Ga or In atoms occupying the 4a, 4c and 8f Wyckoff positions. The only intermediate composition we successfully synthesized was YbNiGa$_{3}$In, with a phase purity of 93\%. Refinement of YbNiGa$_{3}$In suggests that indium has a strong site preference, and almost exclusively occupies the 4a site. (see figure S1) Once this site is fully occupied, the pseudobinary alloy range appears to truncate, evidenced by nominal compositions of YbNiGaIn$_{3}$ and YbNiGa$_{2}$In$_{2}$ not yielding a significant phase fraction of the desired 114-phase.
Previous reports suggest YbNiAl$_{4}$ can be grown via similar methods, but attempts via (1) arc-melting (2) tetra-arc melting (3) induction melting and (4) Al-flux growth with a variety of starting compositions and annealing procedures failed to provide specimens of satisfactory quality. \cite{YNAgrowth} In the cases of (1-2), Yb$_{3}$Ni$_{5}$Al$_{19}$ was the dominant impurity phase, typically 15-25\%, with less than 3\% YbAl$_{3}$. Annealing did not improve phase purity, and in some instances increased the 3-5-19 phase. Attempts to grow YbNiAl$_{4}$ via (3) induction melting as well as attempts to anneal ingots from (1) in Yb-atmosphere resulted in a series of new peaks appearing (see figures S2 and S3), which are comparable in intensity to the 114-peaks and have not been successfully indexed. In the case of (4) we grew only single crystals of Yb$_{3}$Ni$_{5}$Al$_{19}$, consistent with previous results. \cite{3519}
][
	Ambient pressure XRD patterns were collected with a standard D8 diffractometer and were used to determine the lattice parameters and atomic positions. Refinement of the ambient pressure data suggests all sites are within 2\% of being fully occupied. Pressure-dependent XRD studies were performed at sector 16-BMD of the Advanced Photon Source (APS) using a 30 keV x-ray beam. Powdered samples were loaded into a diamond anvil cell (DAC) with a rhenium gasket and pressurized with neon. The pressure was determined via copper powder mixed with the samples and confirmed with ruby-spectroscopy at select pressures.\cite{ruby} The XRD patterns were collected via an area detector and converted to powder patterns using fit2d. \cite{Fit2D} A CeO$_{2}$ sample was used as a calibrant to determine the instrument parameters used in the fitting, which was performed using GSAS-I. \cite{GSAS, EXPGUI} The instrument parameters and atomic positions were then held constant for all subsequent refinements; only lattice parameters, broadening due to strain, and preferred orientation parameters were allowed to vary under pressure. Because of peak broadening that occurs under pressure, some peaks merge  and become difficult to index under pressure. If merging peaks prevent an adequate fit, they are removed for the pressure space in which they overlap. In instances where including and excluding the peaks over various pressure ranges causes inconsistencies in the lattice parameters, they are removed for all XRD spectra. 

	XAS at the ytterbium L-III edge was performed at sector 16-IDD of the APS using partial fluorescence yield (PFY) from the L-alpha emission line. The incident energy was scanned with a Si (111) fixed-exit, double-crystal monochromator and the L-alpha emission was recorded using three Si (400) analyzers. Samples were loaded into a DAC with a beryllium gasket for XAS measurements and mineral oil was used as a pressure transmitting medium. Pressure was measured via ruby-fluorescence spectroscopy.  For both diffraction and spectroscopy measurements, pressure was increased using a gas-driven membrane. Resonant X-ray Emission Spectroscopy (RXES) was fit using the Kramers-Heisenberg formula for photon-atom scattering. \cite{RXESfit1, RXESfit2} The Yb-volume was found by calculating the voronoi cell, i.e. the volume closer to one atom than any other. For this calculation, the  atomic positions are held constant under pressure and the atoms were weighted by covalent radii.

\begin{figure}[b]
	\centering
	\includegraphics[width=\linewidth]{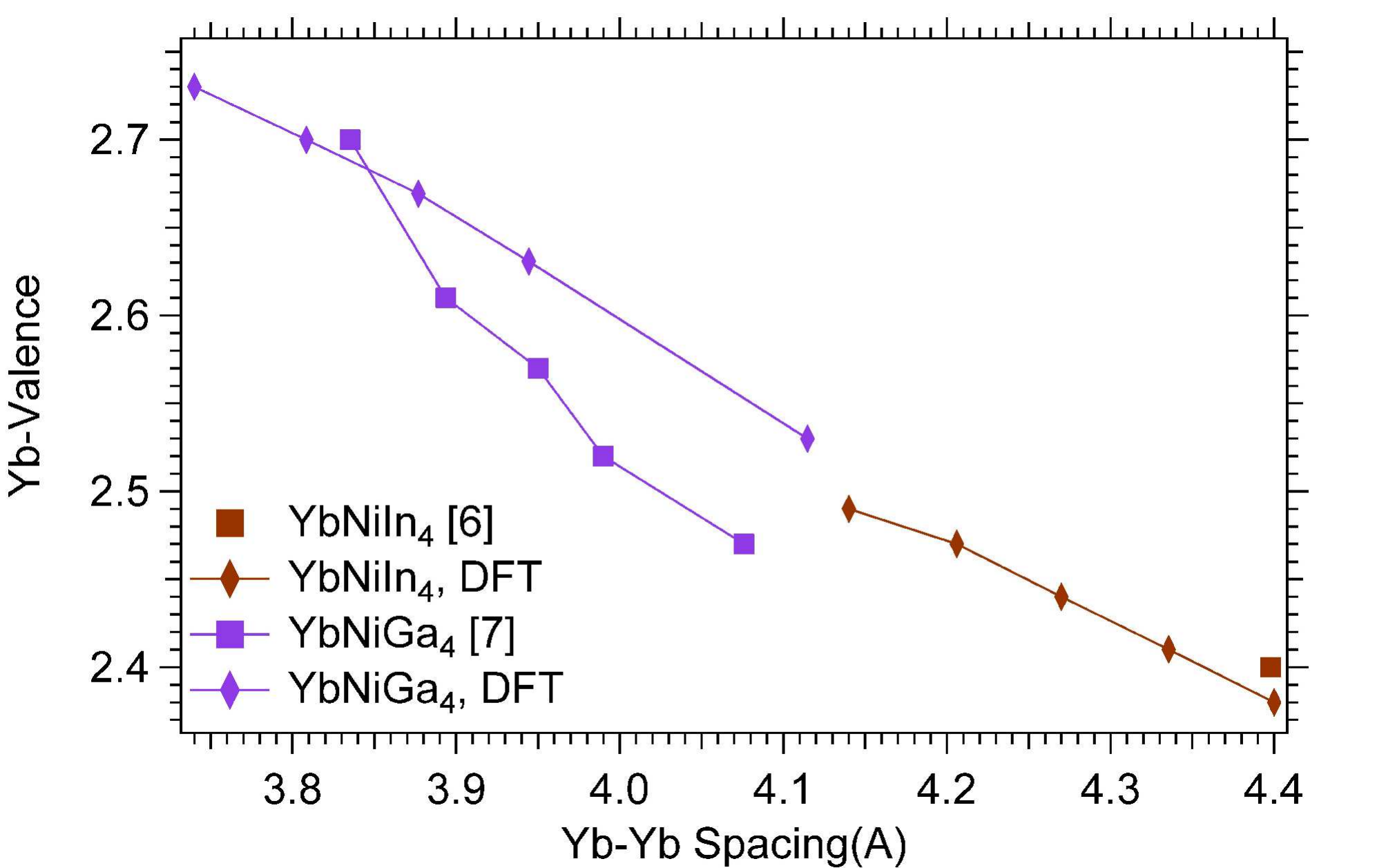}
	\caption{(color online) Summary of DFT calculations and previously published work. DFT reproduces the general behavior, but implies a stronger dependence on Yb-Yb spacing. }
	\label{fig:DFT}
\end{figure}

\begin{figure*}[htbp]
	\centering
	\includegraphics[width=\linewidth]{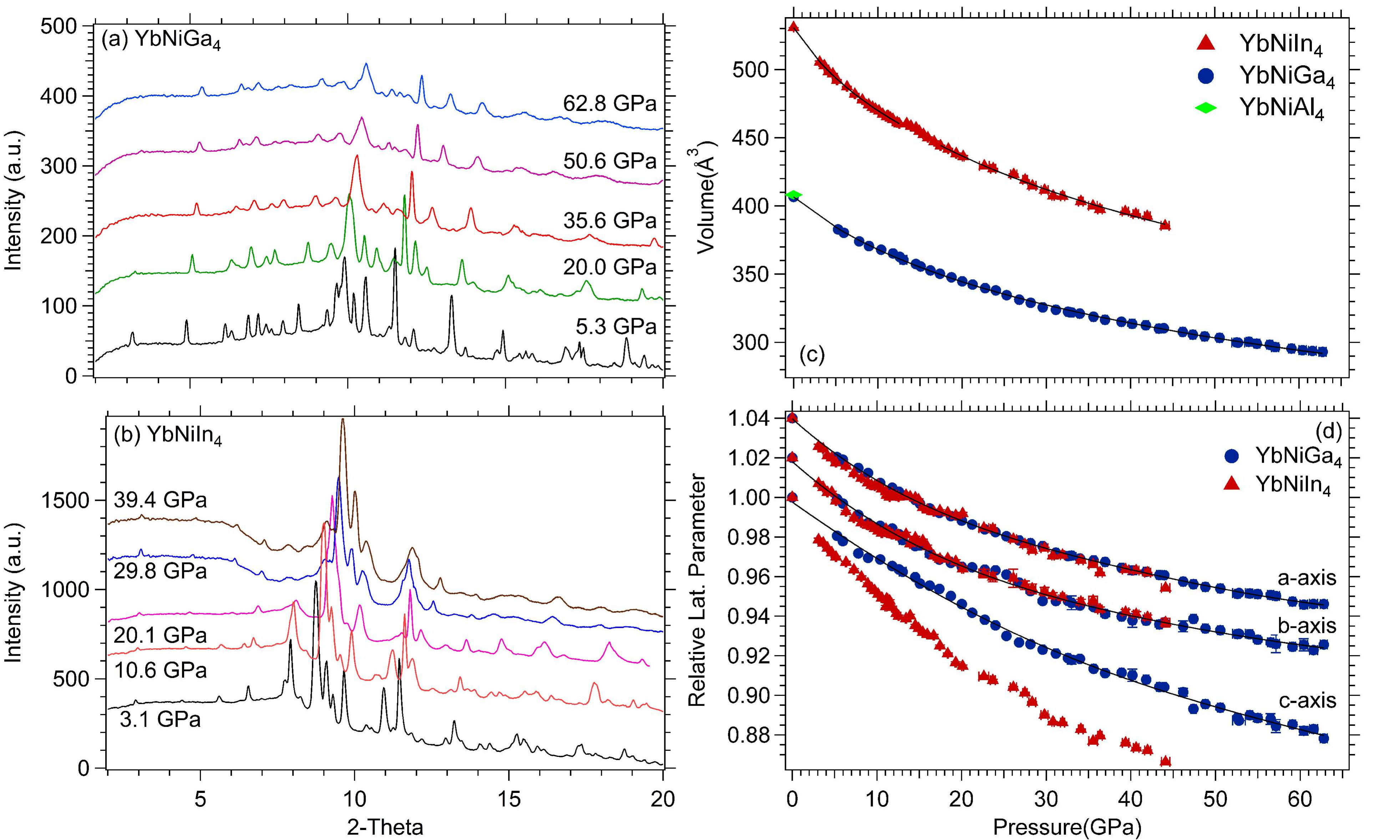}
	\caption{(color online) PXRD spectra at select pressures for (a) YbNiGa$_{4}$ and (b) YbNiIn$_{4}$. In (b), the spectra shown for 29.8 GPa and 39.4 GPa were acquired for a separate measurement than those presented for pressures below, which accounts for some of the difference in intensities of the sample peaks. (c) Volume contraction as a function of pressure. (d) Contraction of lattice parameters as a function of pressure. The a- and b-axes appear to contract similarly among the compounds, but the c-axis displays significantly different behavior. The uncertainties are taken from the uncertainty in GSAS fitting and are generally smaller than the markers. The solid lines in (c) represent a fit to the Birch-Murnaghan equation of state. The solid lines in (d) are guides to the eye, and the a- and b-axes are offset by 0.04 and 0.02, respectively.}
\label{fig:structure}
\end{figure*}

	All calculations are performed within the framework of DFT and similar to a recent study on the rare-earth elemental metals. \cite{DFT} The necessary assumption for the unknown electron exchange and correlation functional is chosen to be that of generalized gradient approximation. The implementation is done for a full-potential linear muffin-tin orbitals (FPLMTO) method. \cite{FPLMTO} The orbital polarization (OP) is included in the FPLMTO as a parameter-free scheme where an energy term proportional to the square of the orbital moment is added to the total energy functional to account for intra-atomic interactions. It is an approximate method that is analogous to the mean-field approximation for the spin-polarization energy. An f$_n$ atomic configuration involves intra-atomic interactions such as (vector model) \textbf{s$_i$ $\cdot$ s$_j$} and \textbf{l$_i$ $\cdot$ l$_j$} (electron i spin and angular momenta). Here, we replace the energy associated with the angular momenta, \\-$\dfrac{1}{2}\sum$\textbf{l$_i$ $\cdot$ l$_j$}, with a mean-field expression, -$\dfrac{1}{2}$($\sum$l$^z_i$)($\sum$l$^z_j$ ) (z component of vector \textbf{l}). This term is proportional to L$^2$ in analogy with the Stoner energy for spin polarization, -$\dfrac{1}{2}$($\sum$s$^z_i$)($\sum$s$^z_j$), that is proportional to M$^2$. L and M are here the total orbital and spin moments, respectively. In the OP scheme this then provides for a one-electron eigenvalue shift proportional to -Lm$_l$ (for each state m$_l$) that enhances the orbital polarization over the spin-orbit coupling only case. The connection between OP and the LDA + U methodologies was discussed recently. \cite{LDA} One distinct advantage with the OP scheme over the LDA + U method is that the former does not depend on a parameter whose pressure dependence is unknown.
	
\begin{figure}[b]
	\centering
	\includegraphics[width=\linewidth]{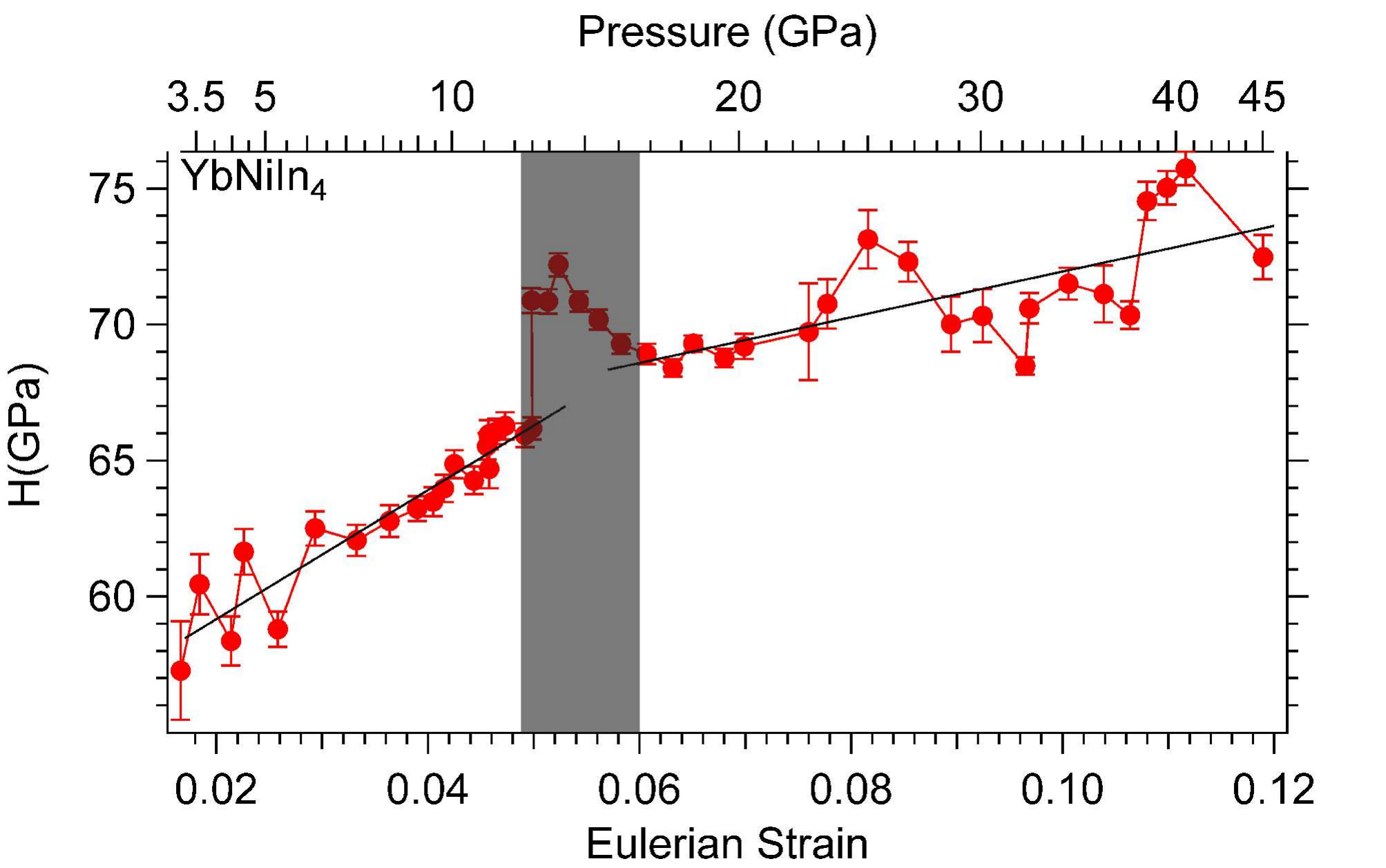}
	\caption{(color online) Reduced pressure, H, plotted against Eulerian strain for YbNiIn$_{4}$.  YbNiIn$_{4}$ displays a sharp spike as well as a change in slope just above a Eulerian strain of 0.05, suggestive of an ETT. The solid black lines represent linear fits below a Eulerian strain of 0.05 and above 0.06. The shaded area represents the transition region.}
	\label{fig:Strain}
\end{figure}
	
\section{\label{sec:level1}Results}

\subsection{\label{sec:level2}DFT}
The results of the DFT calculations are shown in figure ~\ref{fig:DFT} and compared to previously published work. DFT predicts a continuous increase in Yb-valence with decreasing interatomic distance and reproduces the general trend of the previously published experimental data. The DFT calculations, however, suggest a stronger dependence on Yb-Yb spacing than observed, thus predicting a smaller valence for YbNiIn$_{4}$, but predicting a larger than observed valence for YbNiGa$_{4}$, implying that electron correlation effects beyond what is included in the present DFT calculations may be responsible.

\subsection{\label{sec:level2}Structural Studies}

XRD measurements were performed on YbNiGa$_{4}$ and YbNiIn$_{4}$ up to a pressure of 63 GPa and 45 GPa, respectively. Figure ~\ref{fig:structure} shows that YbNiGa$_{4}$ contracts without any sign of a structural transition, and can be well described by the Birch-Murnaghan equation of state (BM-EOS) with B=76.7 GPa and B'=5.5. ~\cite{BM} YbNiIn$_{4}$ shows a plateau in volume between 12 GPa and 14 GPa, which is caused by plateaus in the a- and b-axes in this pressure range (figure 2d). Below 12.5 GPa, the BM-EOS yields B=54.2 GPa and B'=7.0, and the high-pressure region above 17 GPa yields B=63 GPa and B'=5.0, values that more closely resemble those of YbNiGa$_{4}$. To better determine the origin of the plateau near 13 GPa, we have performed a linearization of the BM-EOS as described in reference \cite{strain} and plot the resulting reduced pressure, H, vs the Eulerian strain, f$_{E}$, in figure ~\ref{fig:Strain}. Plotting the reduced pressure vs Eulerian strain should be linear for any stable compound, while a change in slope may be indicative of an electronic topological transition (ETT). As shown in fig. \ref{fig:Strain}, YbNiIn$_{4}$ shows a sudden spike in the reduced pressure at a Eulerian strain of 0.05 (corresponding to 12.5 GPa), which is accompanied by a change in slope, which may indicate the presence of an ETT.

\subsection{\label{sec:level2}Valence Measurements}
\subsubsection{X-ray Absorption Spectroscopy}

\begin{figure}[hbtp]
\centering
\includegraphics[width=\linewidth]{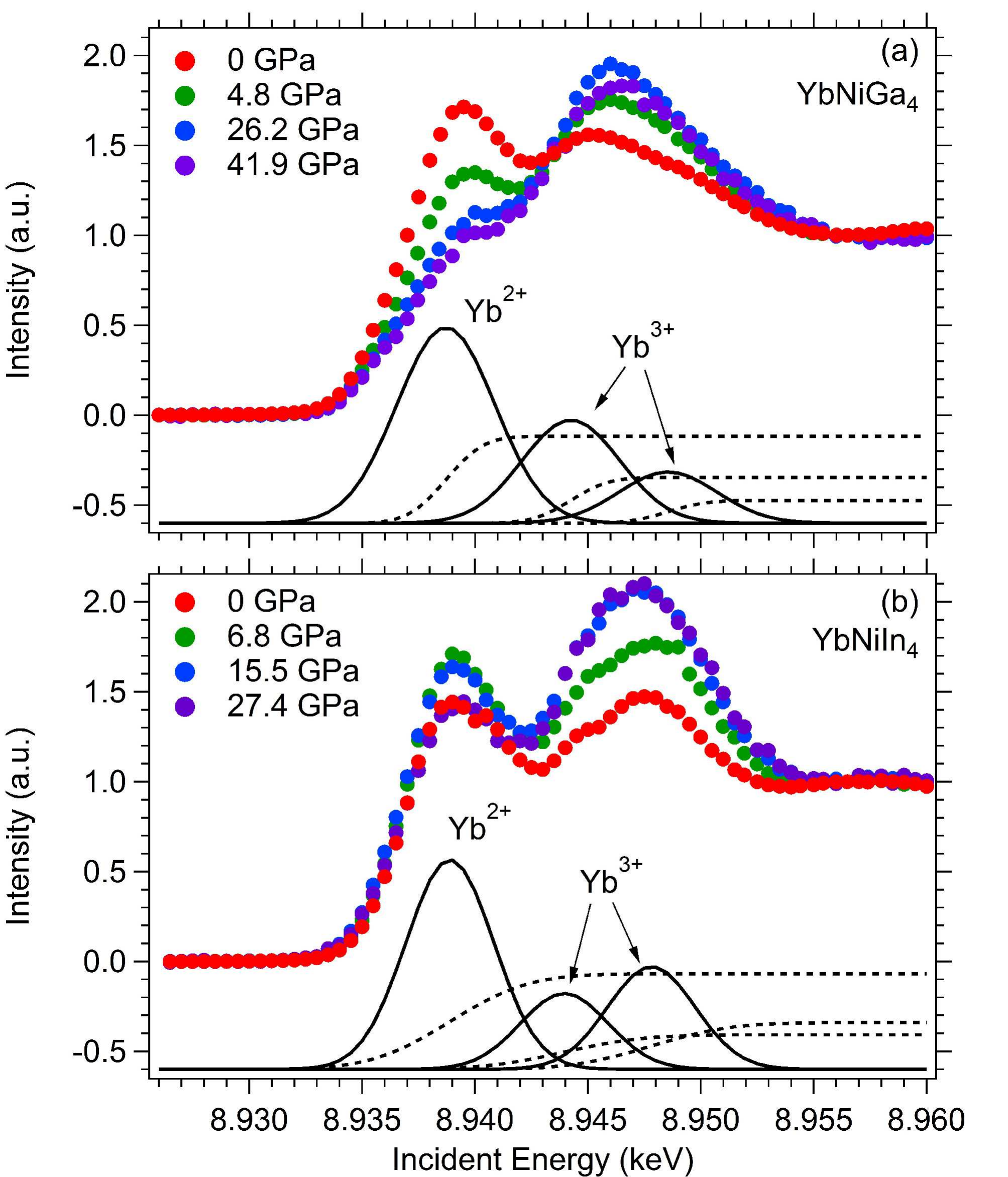}
\caption{(color online) X-ray absorption spectra at select pressures and the fit functions for (a) YbNiGa$_{4}$ and (b) YbNiIn$_{4}$. The spectra are normalized to an edge jump of unity. The solid lines of the fit correspond to the gaussian functions associated with the valence peaks and the dashed lines are their respective error functions accounting for entering the fluorescent region. The Yb$^{2+}$ and Yb$^{3+}$ peaks are indicated. With increasing pressure, the Yb$^{3+}$ peak gains intensity, while the Yb$^{2+}$ peak loses intensity. }
\label{fig:XAS}
\end{figure}

XAS is sensitive to the valence of the Yb-ions because the 4f-states (4f$^{13}$ and 4f$^{14}$) each experience different screening. Each valence state (Yb$^{3+}$ and Yb$^{2+}$) will result in a distinct absorption peak in the XAS spectra, which are separated by approximately 8-12 eV. By calculating the weighted average of the peak intensities, the effective Yb-valence can be determined. The XAS spectra can be fit by describing each valence state with a Gaussian and error function. As reported in several papers studying Yb-valence in other materials, we observed a splitting of the Yb$^{3+}$ peak, which is likely due to the crystal field splitting of the unoccupied 5d-states.\cite{split1,split2,split3,split4} Previous work measuring the valence of YbNiGa$_{4}$ and YbNiIn$_{4}$ was performed in transmission mode and lacked the resolution to fit both Yb$^{3+}$ peaks, resulting in a minor difference in the determined valence and pressure dependence thereof compared to the work reported herein.\cite{YNI,YNG} Figure ~\ref{fig:XAS} shows the details of our fit for the ambient pressure data, as well as several XAS spectra at select pressures. 

\begin{figure}[hbtp]
\centering
\includegraphics[width=\linewidth]{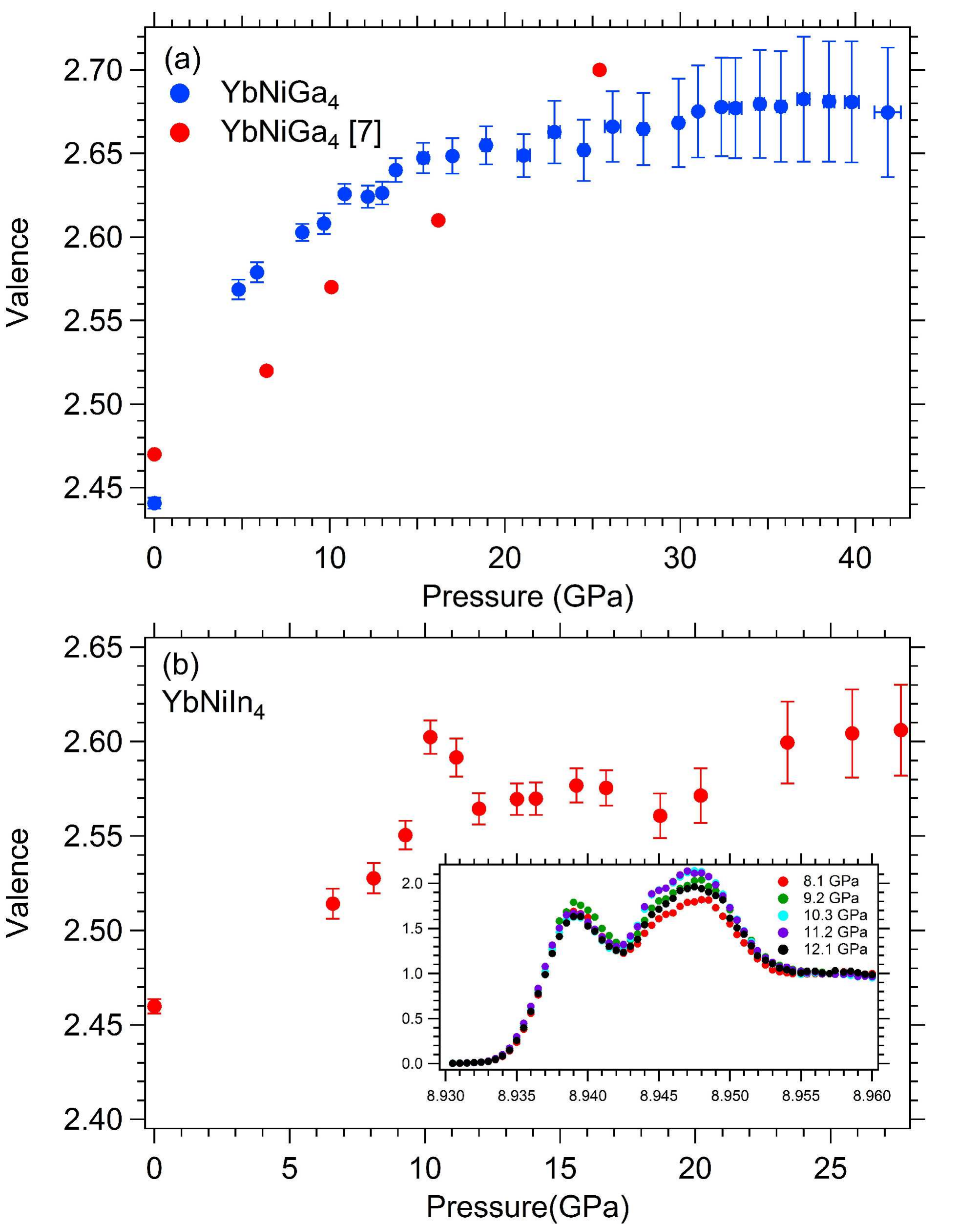}
\caption{(color online) Yb-valence as a function of pressure determined via XAS for (a) YbNiGa$_{4}$ and (b) YbNiIn$_{4}$. Inset: XAS spectra of YbNiIn$_{4}$ around 10 GPa. The valence for YbNiGa$_{4}$ increases up to about 25 GPa, above which the valence appears to be saturated. For YbNiIn$_{4}$, the valence reveals a peak near 10 GPa, consistent with the ETT proposed from structural results. Aside from this peak, the valence increases steadily up to the maximum pressure of P=27 GPa, though the highest measured pressure points suggest the valence may be reaching saturation. The uncertainties were calculated from weighted fitting in Igor.}
\label{fig:Ybval}
\end{figure}

For YbNiGa$_{4}$, there exists a clear decrease in the intensity of the Yb$^{2+}$ peak and increase in intensity of the Yb$^{3+}$ peak up to a pressure of ~25 GPa. Above 25 GPa, the ratio of absorption peak to fluorescence decreases, but the valence remains largely unchanged. For YbNiIn$_{4}$, the ambient pressure measurement reveals a larger contribution from the fluorescent region than the subsequent pressure measurements, resulting in an apparent increase of both the Yb$^{2+}$ and Yb$^{3+}$ peaks from ambient to 6.8 GPa. Nonetheless, the ratio of amplitudes of these valence peaks results in an increase in valence with pressure, following the trend observed for all the measured pressures. Summarizing the valence determination via XAS and adding previous valence determinations for YbNiGa$_{4}$ yields figure ~\ref{fig:Ybval}. \cite{YNG} 

The Yb-valence of YbNiGa$_{4}$ increases up to about 20-25 GPa, at which point the valence saturates at n=2.68. The Yb-valence in YbNiIn$_{4}$ may be approaching saturation at the highest measured pressures, but there is also a peak in valence close to 10 GPa. This peak is likely another manifestation of the ETT which was observed in the structural measurements, and could be the result of the changing electronic density of states near the ETT.

\subsubsection{Resonant X-ray Emission Spectroscopy}
\begin{figure}[hbtp]
\centering
\includegraphics[scale=.5]{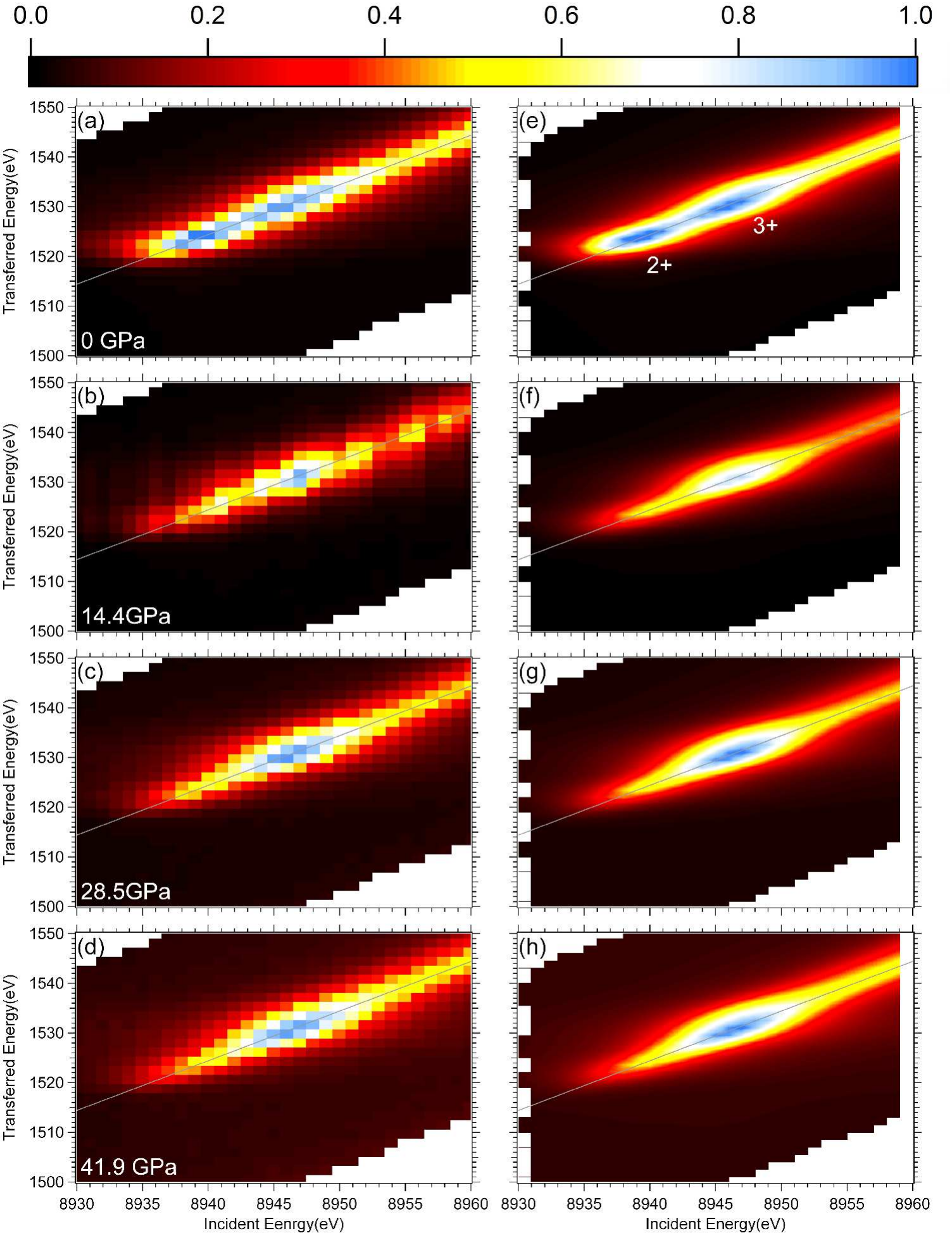}
\caption{(color online) (a-d) RXES spectra and (e-h) the corresponding fit for YbNiGa$_{4}$. With increasing pressure, the low energy 2+ peak decreases in intensity. The gray lines correspond to the XAS-PFY scans described above. Intensities are normalized to the maximum intensity of the 3+ peak of the experimental data.}
\label{fig:RXES}
\end{figure}

RXES is a powerful tool for fully describing the valence state of a given material, which scans the emission energy in addition to the incident energy. Converting the emitted energy to transferred energy and combining this into a single plot results in the RXES spectra shown in figure ~\ref{fig:RXES} for YbNiGa$_{4}$ and figure S4 for the single pressure measured for YbNiIn$_{4}$. As in the case of PFY measurements, the amplitudes of the absorption peaks allow for determination of the valence. 

YbNiGa$_{4}$ begins with a rather broad peak due to the overlap of the three distinct contributions of the measured valence peaks, but with increasing pressure the valence state is shifted away from the Yb$^{2+}$ and towards the Yb$^{3+}$ state. This results in only a weak Yb$^{2+}$ structure remaining at 42 GPa. The overall trend of the valence determined from RXES and XAS is the same for YbNiGa$_{4}$. The valence peaks in YbNiIn$_{4}$ have a larger separation resulting in more distinct peaks and agreeing with the valence determined via XAS. 

\section{\label{sec:level1}Discussion}

While both the structural and spectroscopic data of YbNiIn$_{4}$ are suggestive of an ETT, it is important to note that previous work measuring ETTs have not observed a peak in reduced pressure, or conversely, did not show a plateau in volume near the ETT. \cite{strain,Bi2Te3,Sb2Te3} We speculate that this is due to the dual nature of the Yb 4f-electrons, which display both local and itinerant character in these intermediate valence states observed in YbNiIn$_{4}$. Previous pressure-induced ETT have been identified in weakly correlated itinerant systems, implying compressibility changes arising only from the bonding changes driven by subtle changes near the Fermi level. In the case of YbNiIn$_{4}$, the dual nature of the 4f-electron subsystem yields consequences not only for the electronic structure near the Fermi level, as with the itinerant systems, but also for the local, core-like states, which have intendent ramifications to the ionic volume of the Yb atoms independent of the physics at the Fermi level. The physics that drives the dual nature of the 4f-electrons in YbNiIn$_{4}$ inherently couples the local part of the wavefunction to the ETT, which may be expected to yield more pronounced effects on compressibility and volume than typically seen in weakly correlated, itinerant systems.

\begin{figure}[hbtp]
\centering
\includegraphics[width=\linewidth]{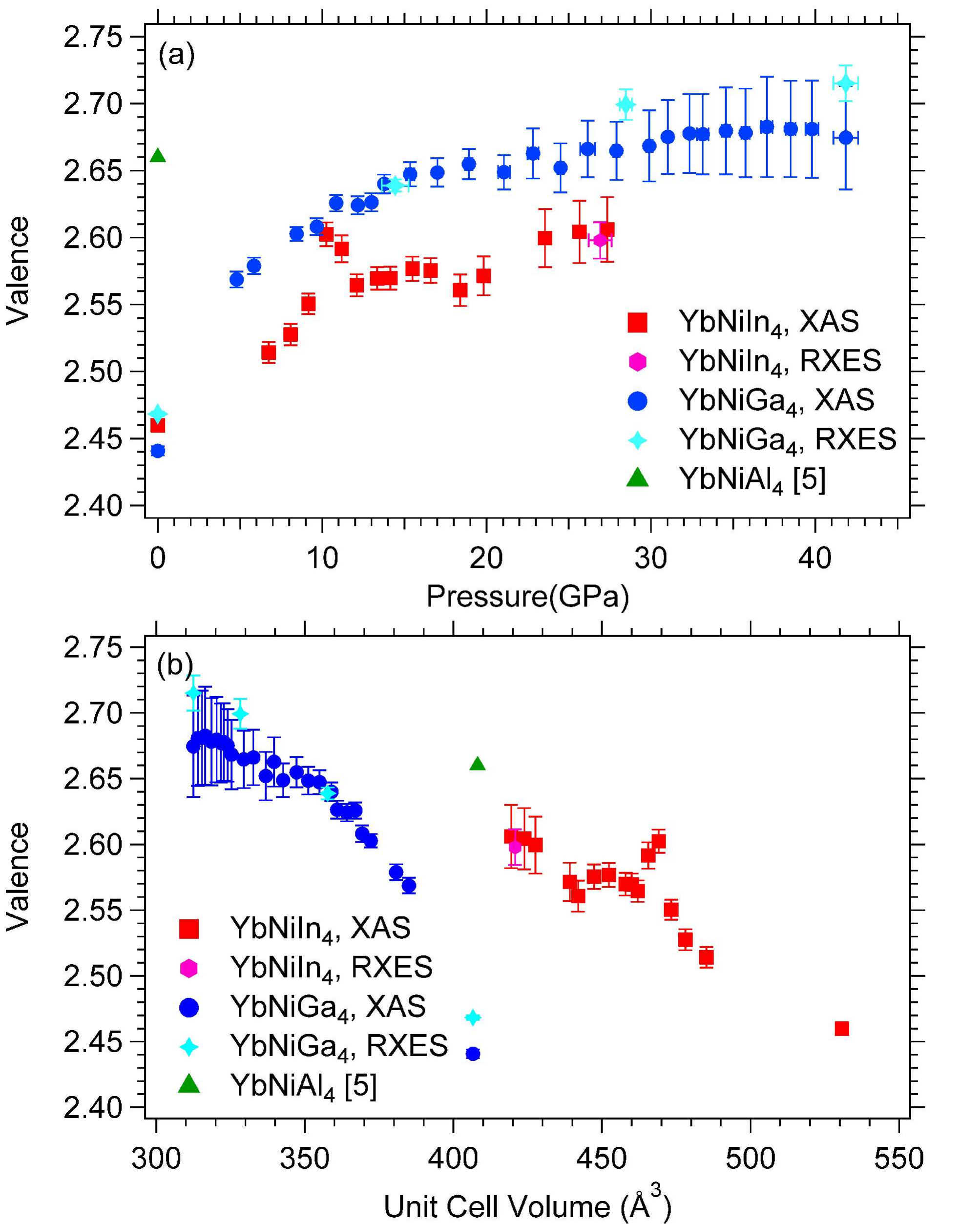}
\caption{(color online) Summary of valence measurements of this system plotted against (a) pressure and (b) unit cell volume. With increasing pressure, the Yb-valence increases, though in YbNiGa$_{4}$ the Yb-valence appears to reach saturation above ~25 GPa. Unit cell volume is insufficient to fully describe the Yb-114 system.}
\label{fig:Press_Vol}
\end{figure}

Figure ~\ref{fig:Press_Vol}a summarizes the valence determined via each of the described methods, and includes the valence determined for YbNiAl$_{4}$ from reference 5. YbNiGa$_{4}$ appears to reach saturation at n=2.68 above P=25 GPa and surpasses the Yb-valence measured in YbNiAl$_{4}$. Both YbNiGa$_{4}$ and YbNiIn$_{4}$ have comparable Yb-valence at ambient pressure, but in YbNiGa$_{4}$ the valence appears to be more sensitive to pressure. As shown in figure ~\ref{fig:Press_Vol}b, the unit cell volume fails to describe the overlapping region of these materials, though both materials individually reveal the expected trend of increasing valence with decreasing unit cell volume. 

\begin{figure}[hbtp]
\centering
\includegraphics[width=\linewidth]{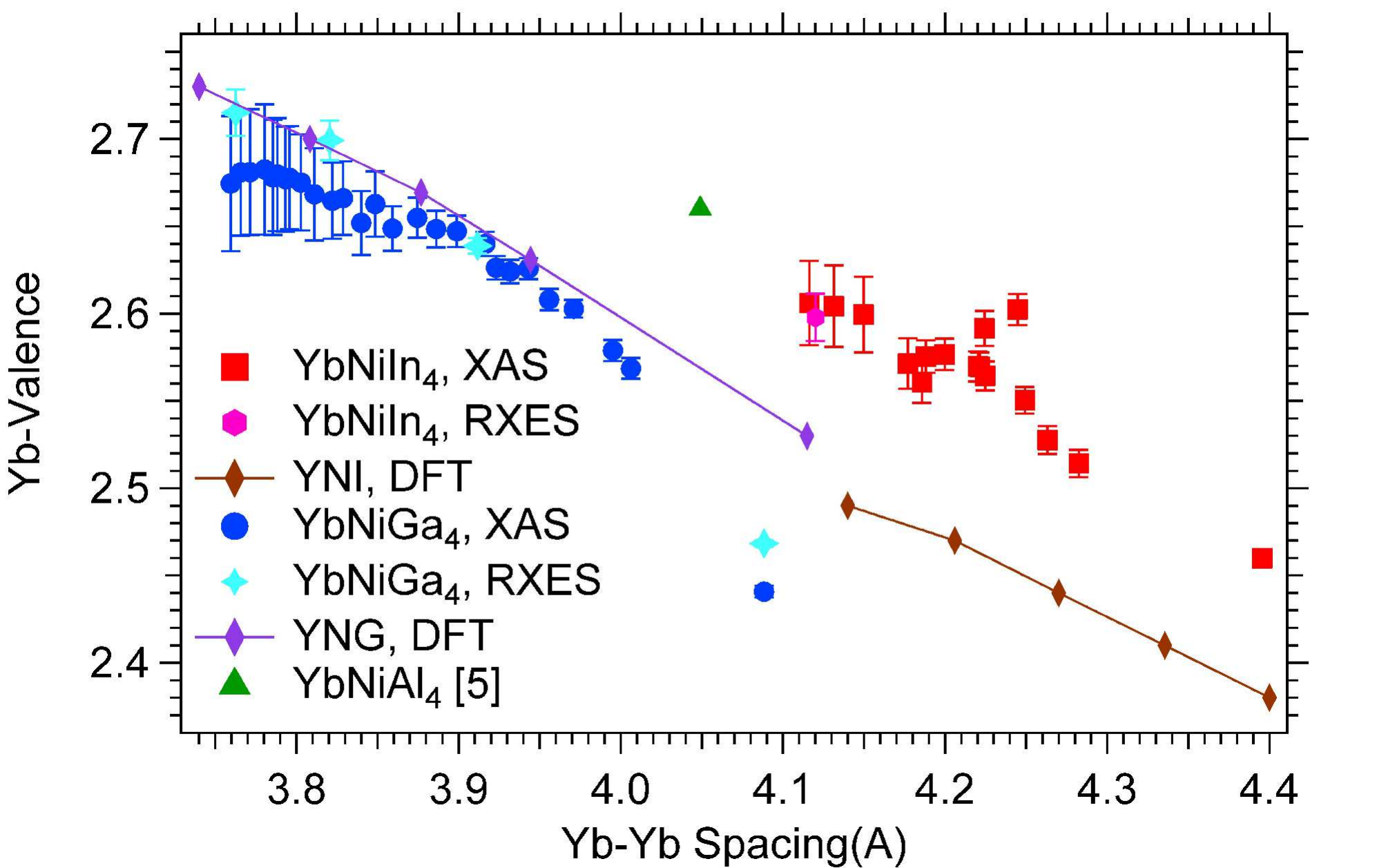}
\caption{(color online) Yb-valence plotted against Yb-Yb spacing and DFT calculations for Yb-valence vs Yb-spacing. DFT does not reproduce the sharp drop when transitioning from YbNiIn$_{4}$ under pressure to YbNiGa$_{4}$ under ambient conditions. }
\label{fig:Ybspacing}
\end{figure}

Figure ~\ref{fig:Ybspacing} shows Yb-valence vs Yb-Yb spacing, which shows similar behavior as unit cell volume. Most of the data for YbNiIn$_{4}$, the high-pressure region of YbNiGa$_{4}$, as well as YbNiAl$_{4}$ at ambient pressure appear to follow a smooth valence vs Yb-Yb spacing curve. However, in the region where these compounds have similar Yb-Yb spacing, the Yb-valence reveals a precipitous drop, indicating that Yb-Yb spacing does not capture the entirety of the underlying physics. Figure ~\ref{fig:Ybspacing} also includes DFT calculations for the Yb-valence in the YbNiGa$_{4}$ and YbNiIn$_{4}$ systems. While the zero-pressure value of valence in YbNiIn$_{4}$ predicted by DFT is lower than the experiments, the pressure-dependent trend of the valence as predicted by DFT is in good agreement with the experimental observations. For YbNiGa$_4$ the behavior is also reproduced well for smaller lattice spacings while the drastic drop at the larger lattice spacings is not predicted by the theory. It is particularly the measured valence at the largest Yb-Yb spacing (4.07 $\AA$) that deviates from theory (2.44 vs 2.54) and the reason is not clear. We speculate that electron correlation effects beyond what is included in the present DFT calculations may be the cause. 

\begin{figure}[hbtp]
\centering
\includegraphics[width=\linewidth]{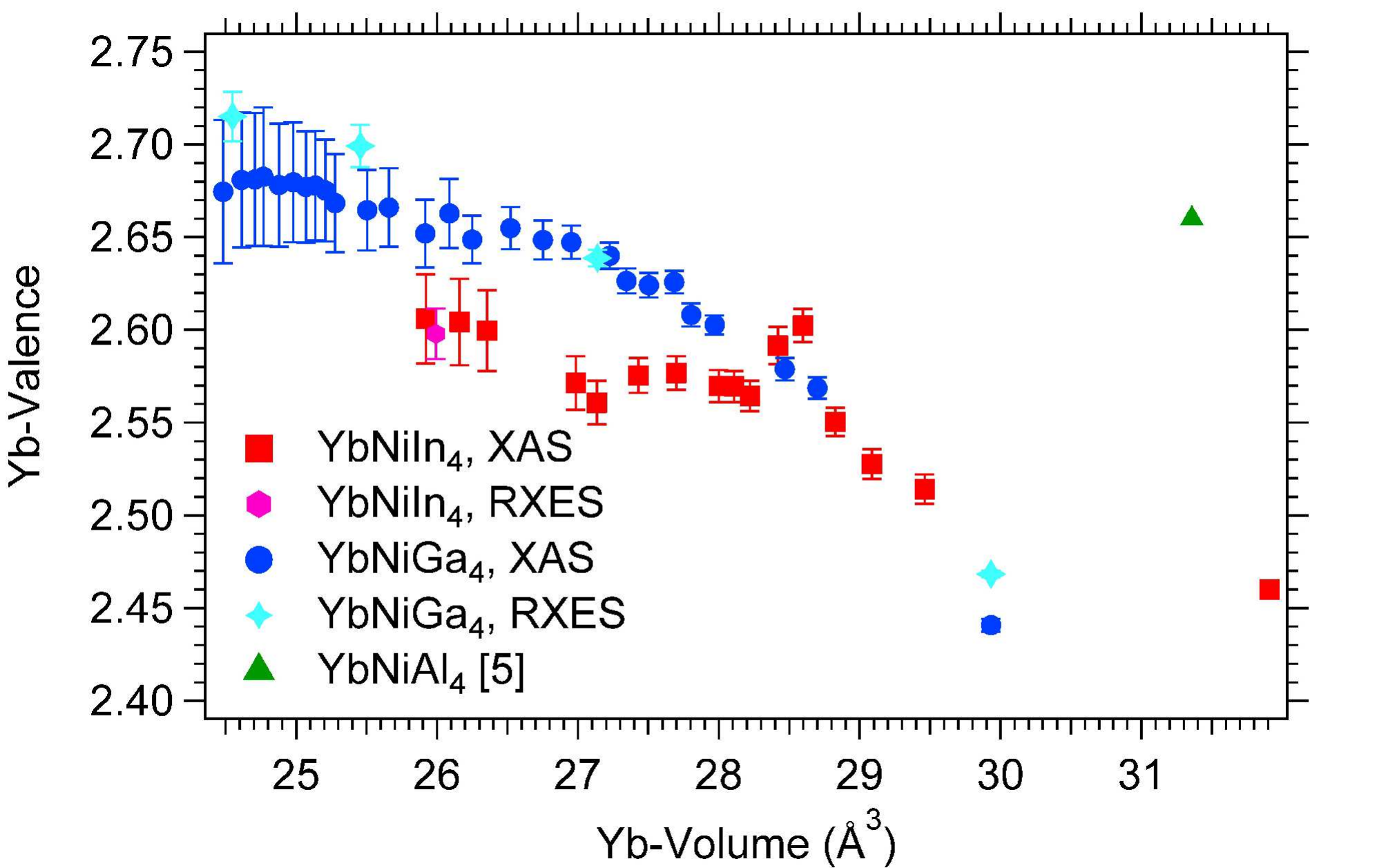}
\caption{(color online) Yb-valence plotted against Yb-volume. There is no convincing trend, and YbNiAl$_{4}$ does not fit within this framework.  }
\label{fig:Ybvol}
\end{figure}

As an attempt to account for the effect of substituting In and Ga, we calculated the Yb-volume, i.e. the space available to the Yb-atoms, for each measured pressure. The results are shown in figure ~\ref{fig:Ybvol}. Consideration of the Yb-volume does not provide a satisfactory result, and YbNiAl$_{4}$ does not fit into this scheme. In the case of Yb-volume, the In- and Ga-variants are comparable and exhibit similar slopes, but this still does not fully capture the evolution of the Yb-valence. This, combined with the Yb-spacing and unit cell volume data, suggests that structural parameters alone are insufficient to fully describe the valence behavior of this system. This implies that the Yb-valence is also sensitive to the chemical environment, i.e. the hybridization between the Yb 4f- and group IIIb p-states, an effect not captured by DFT calculations.   

\section{\label{sec:level1}Conclusion}
In summary, the Yb-valence in the Yb-114 system can be readily modified by pressure, but that valence is not simply described by nearest-neighbor bond distances. By using partial fluorescence yield measurements, we have improved the resolution of the valence determination in YbNiGa$_{4}$, which reveals a steady increase in valence from n=2.44 up to n=2.68 near P=25 GPa, saturating shortly thereafter. The Yb-valence of YbNiIn$_{4}$ shows similar overall behavior, but we have also observed a sharp valence enhancement in YbNiIn$_{4}$ just above 10 GPa. This peak coincides with a plateau in volume, which we speculate is the result of an electronic topological transition. The Yb-valence is most closely related to the Yb-Yb spacing in this structure,  though this parameter is insufficient to describe the valence across the entirety of the YbNi(Ga,In)$_{4}$ system. The hybridization resulting from the Yb-In, Yb-Ga, and Yb-Al bonds appears to be dependent on atomic species and not just the natural bond lengths set by ionic sizes. 

\section{\label{sec:level1}Acknowledgements}
This work was performed under LDRD (Tracking Code 18-SI-001) and under the auspices of the US Department of Energy by Lawrence Livermore National Laboratory (LLNL) under Contract No. DE-AC52- 07NA27344. Part of the funding was provided through the LLNL Livermore Graduate Scholar Program. HPCAT operations are supported by DOE-NNSA under Award No. DE-NA0001974, with partial instrumentation funding by NSF. P. Chow, Y. Xiao, C. Kenney-Benson, R. Ferry and D. Popov acknowledge the support of DOE-BES/DMSE under Award DE-FG02-99ER45775. Beamtime was generously provided through the GUP system and through CDAC. This material is based upon work supported by the National Science Foundation under Grant No. NSF DMR-1609855. R.E.B. and K.H. performed crystal synthesis experiments at the National High Magnetic Field Laboratory (NHMFL), which is supported by National Science Foundation Cooperative Agreements
No. DMR-1157490 and No. DMR-1644779 and the state of Florida. R.E.B. and K.H. were supported in part by the Center for Actinide Science and Technology, an Energy Frontier Research Center funded by the US Department of Energy (DOE), Office of Science, Basic Energy Sciences (BES), under Award No. DE-SC0016568. D.J.C acknowledges the support of the U.S. Department of Energy, Office of Science, Office of Workforce Development for Teachers and Scientists, Office of Science Graduate Student Research program, administered by the Oak Ridge Institute for Science and Education for the DOE under contract no. DE‐SC0014664. J.P. and D.J.C. acknowledge support from the Gordon and Betty Moore Foundation's EPiQS Initiative through grant no. GBMF4419.

\end{document}

%\preprint{APS/123-QED}

\title{Supplemental Information: Pressure dependent intermediate valence behavior in YbNiGa$_{4}$ and YbNiIn$_{4}$}% Force line breaks with \\

\author{Z. E. Brubaker$^{1,2}$, R. L. Stillwell$^{2}$, P. Chow$^{3}$, Y. Xiao$^{3}$, C. Kenney-Benson$^{3}$, R. Ferry$^{3}$, D. Popov$^{3}$, S. B. Donald$^{2}$, P. S{\"o}derlind$^{2}$, D. J. Campbell$^{4}$, J. Paglione$^{4}$, K. Huang$^{5}$, R. E. Baumbach$^{5}$, R. J. Zieve$^{1}$ and J. R. Jeffries$^{2}$\\
$^{1}$ Physics Department, University of California, Davis, California, USA\\
$^{2}$ Lawrence Livermore National Laboratory, Livermore, California 94550, USA\\
$^{3}$ HPCAT, Geophysical Laboratory, Carnegie Institute of Washington, Argonne National  
Laboratory, Argonne, Illinois 60439, USA\\
$^{4}$ Center for Nanophysics and Advanced Materials, Department of Physics, University of Maryland, College Park, Maryland, 20742, USA\\
$^{5}$ National High Magnetic Field Laboratory, Florida State University, Tallahassee, FL 32313, USA\\
}

\date{\today}% It is always \today, today,

%\begin{description}
%\end{description}

\maketitle

%\tableofcontents

\section{\label{sec:level1}Crystal Structure}
Figure 1a shows the crystal structure for YbNiGa$_4$, viewed along the [100] direction. Blue atoms correspond to Yb, green atoms correspond to Ga, grey atoms correspond to Ni, and pink atoms correspond to In. As shown in (b), the In atoms have a strong site preference, and almost exclusively occupy the [000] site. Once this atomic position is fully occupied, the 114-phase no longer forms, evidenced by unsuccessful growth attempts of YbNiGa$_2$In$_2$ and YbNiGaIn$_3$.

\begin{figure}[hbtp]
\centering
\includegraphics[width=0.7\linewidth]{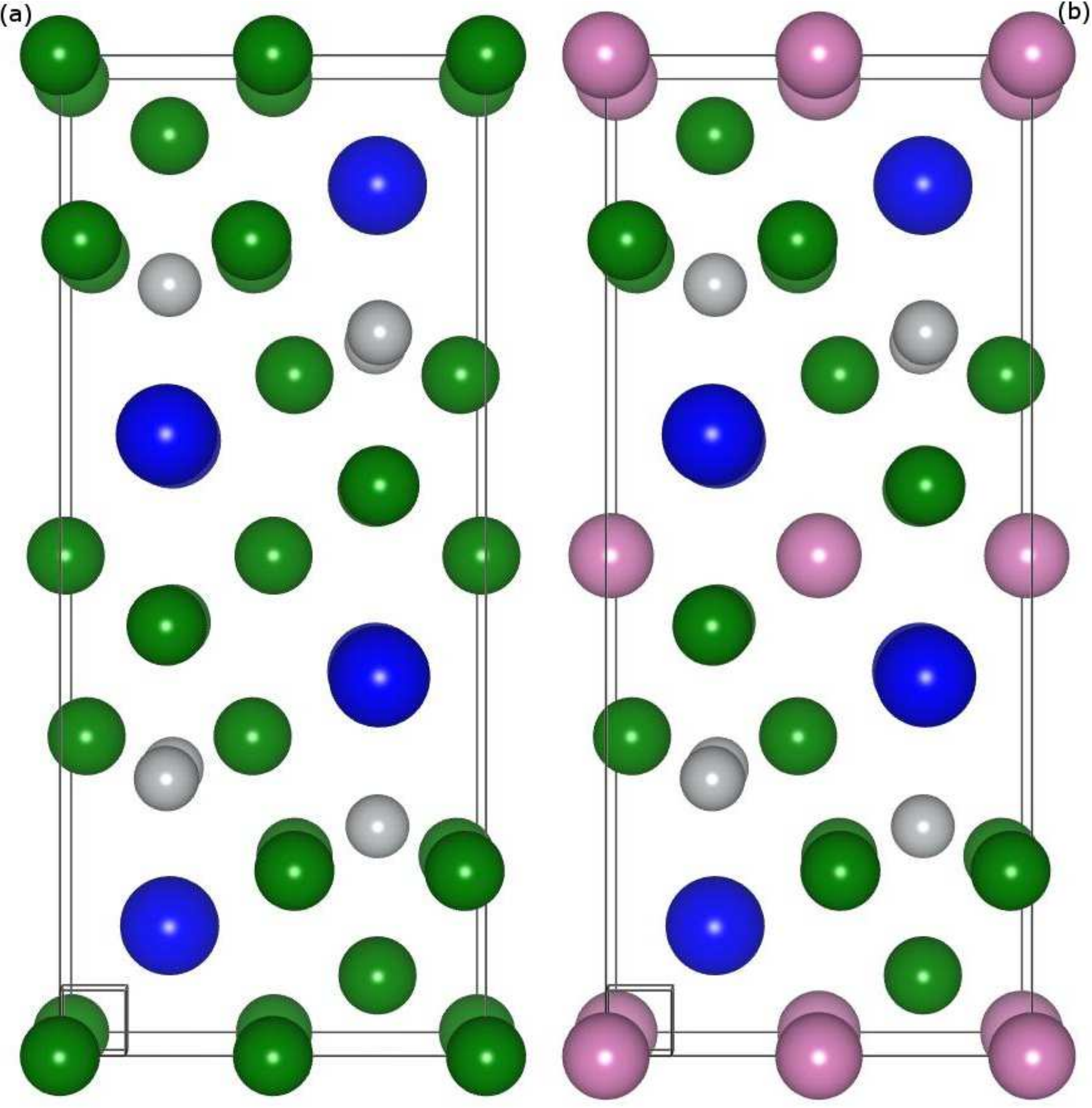}
\caption{Crystal structure of (a) YbNiGa$_4$ and (b) YbNiGa$_3$In viewed along the [100]. }
\end{figure}

\section{\label{sec:level1}X-ray Diffraction for attempted growths of YbNiAl$_{4}$}
Figure 2 shows the XRD spectrum for the growth attempt of YbNiAl$_{4}$ following the same procedure as for YbNiGa$_{4}$ and YbNiIn$_{4}$. The 114-phase is the dominant phase, but there is a large (23\%) contribution from the Yb$_{3}$Ni$_{5}$Al$_{19}$ phase. Figure 3 shows the XRD spectrum for the attempted induction melt, which shows some contribution of the 114-phase, but also reveals several high-intensity peaks which we have not successfully indexed. These same peaks also appeared when annealing the sample from fig. 2 in Yb-atmosphere and when melting additional Yb into the ingot.

\begin{figure}[hbtp]
\centering
\includegraphics[width=\linewidth]{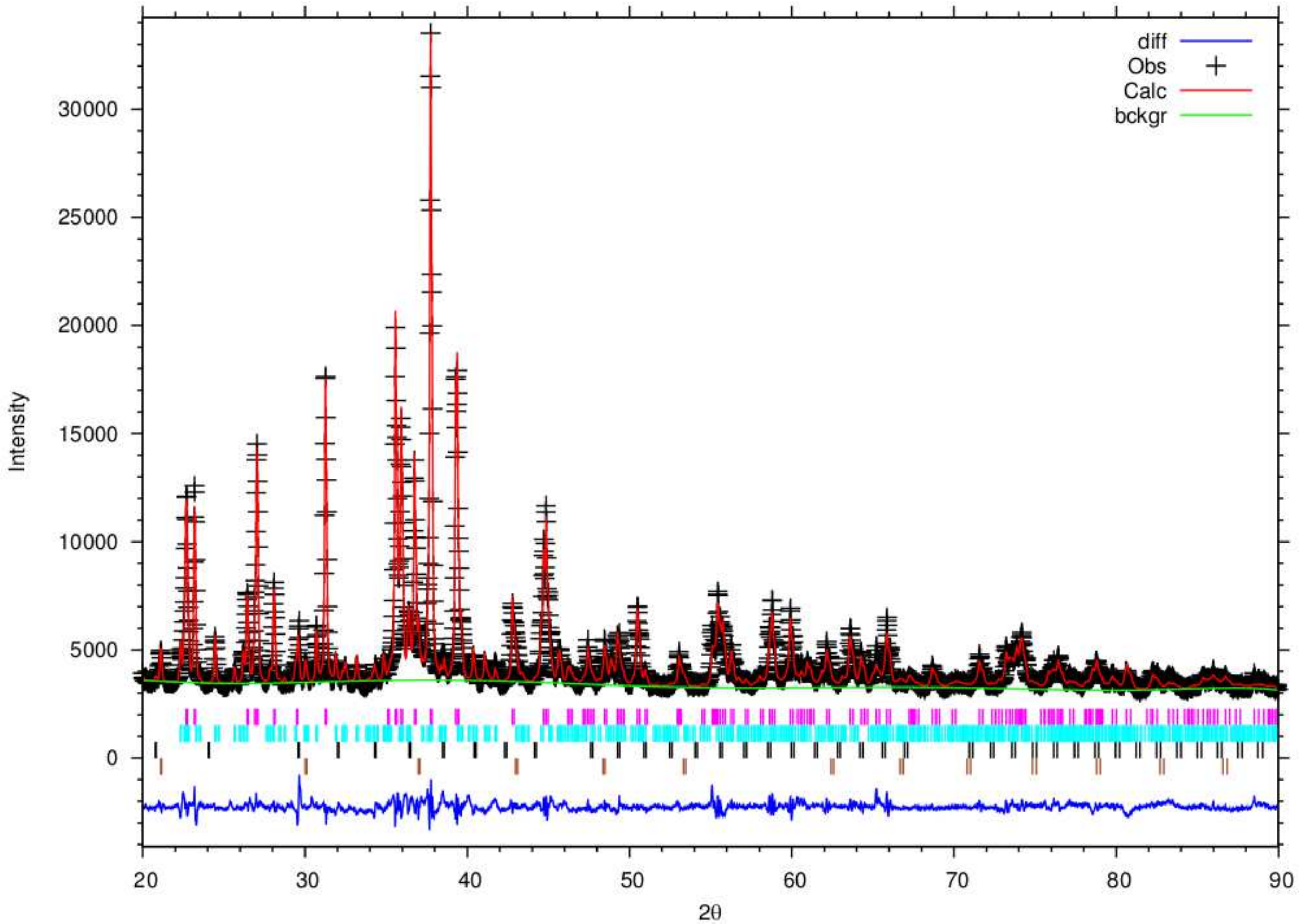}
\caption{XRD spectra for Arc-melted sample. Measured with Cu K-alpha wavelengths. All peaks are indexed, but 3-5-19 makes 23\% contribution. Pink: YbNiAl$_{4}$, 74.1\%. Teal: Yb$_{3}$Ni$_{5}$Al$_{19}$, 23.5\%. Black: Yb$_{2}$O$_{3}$, 0.87\%. Brown: YbAl$_{3}$, 1.46\%.}
\end{figure}

\begin{figure}[hbtp]
\centering
\includegraphics[width=\linewidth]{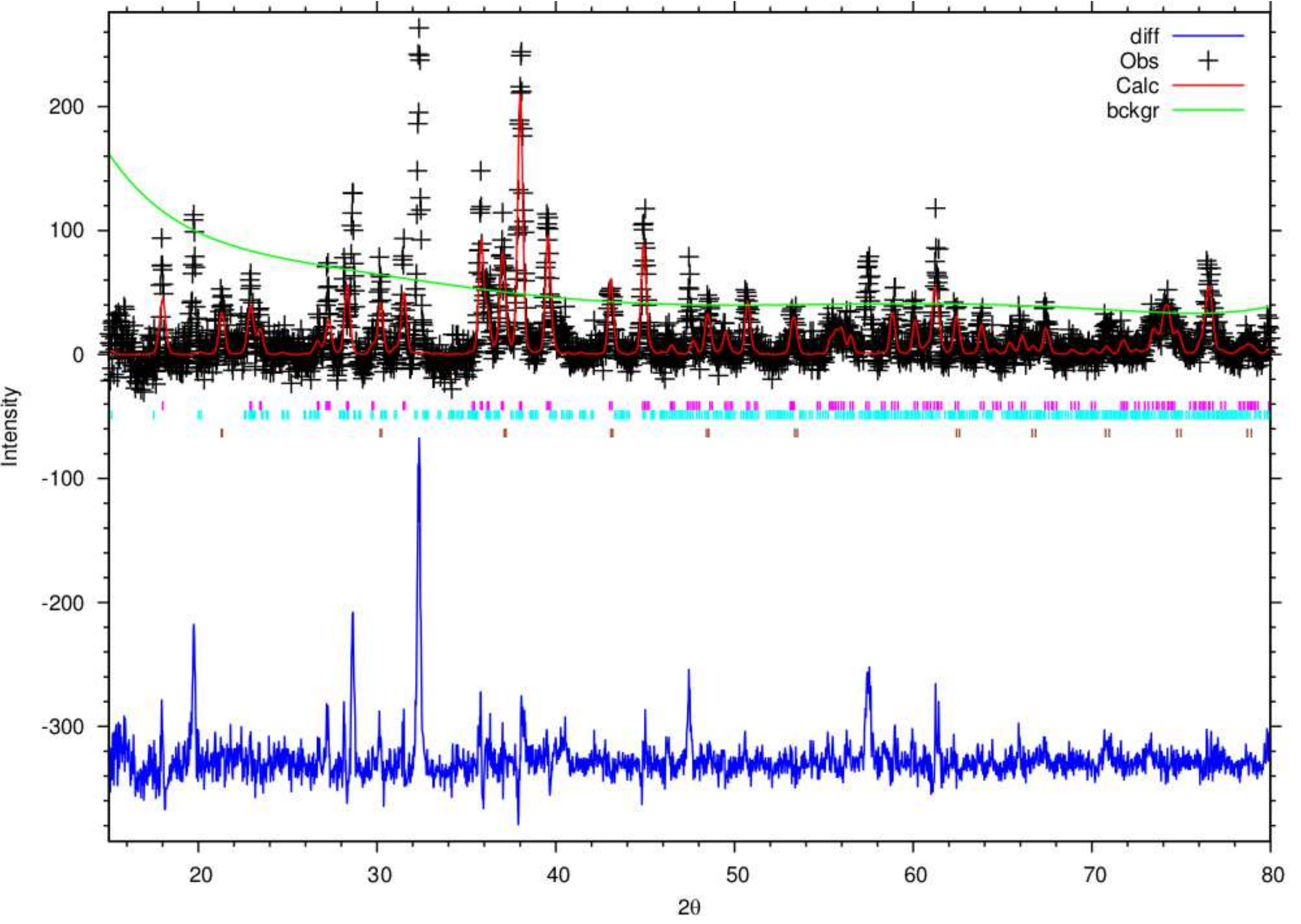}
\caption{XRD spectra for Induction Melting. Measured with Cu K-alpha wavelengths. Unindexed peaks at ~20deg, ~28deg, ~32deg. Pink: YbNiAl$_{4}$, 75.2\%. Teal: Yb$_{3}$Ni$_{5}$Al$_{19}$, 8.9\%. Brown: YbAl$_{3}$, 15.8\%. No sign of Yb$_{2}$O$_{3}$. }
\end{figure}

\section{\label{sec:level1}RXES for YbNiIn$_{4}$}
Figure S4 shows the RXES spectra and fit for YbNiIn4 at 26.9 GPa.
\begin{figure}[hbtp]
\centering
\includegraphics[width=\linewidth]{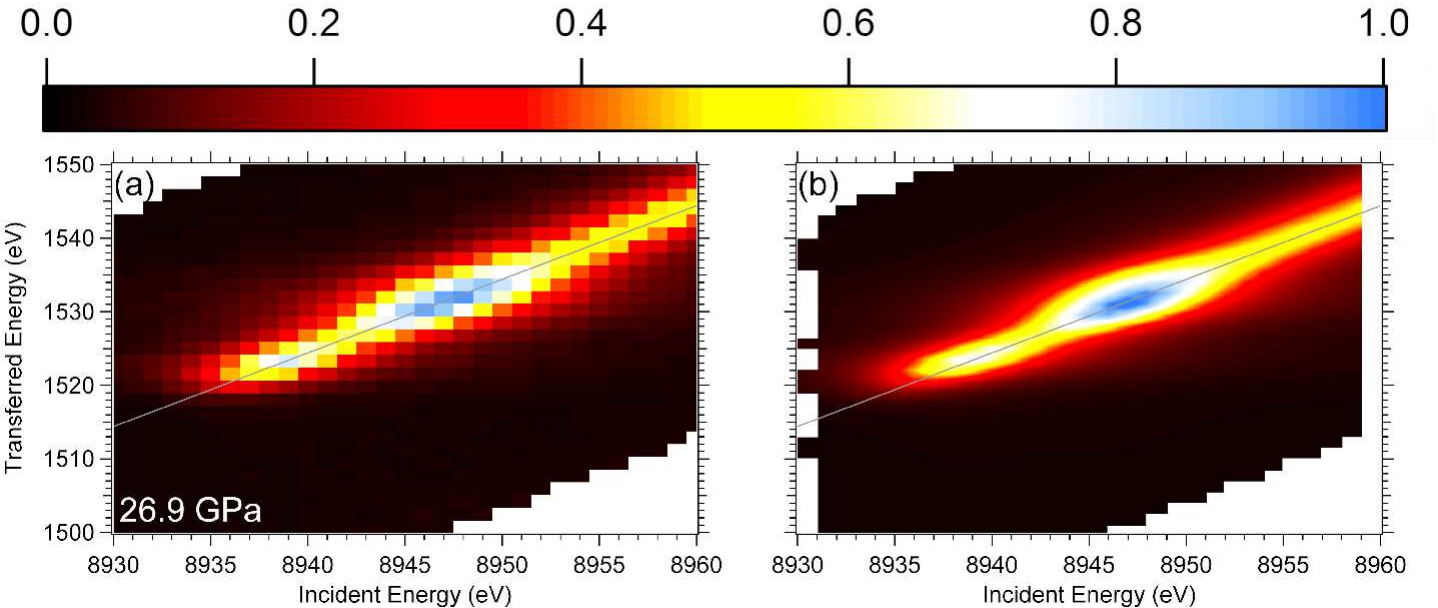}
\caption{(a) RXES spectra and (b) the corresponding fit for YbNiIn$_{4}$. The individual peaks appear to be more clearly separated than in YbNiGa$_{4}$.}
\end{figure}